\documentclass{article}

\usepackage{arxiv}

\usepackage[english]{babel}
\usepackage[utf8]{inputenc}
\usepackage[T1]{fontenc}
\usepackage{comment}
\usepackage{xcolor}
\usepackage{color,soul}
\usepackage{url}            
\usepackage{amsfonts}       
\usepackage{nicefrac}       
\usepackage{microtype}      
\usepackage{lipsum}
\usepackage{tipx}

\usepackage{bm}

\usepackage{amsmath}
\usepackage{amssymb}
\usepackage[version=4]{mhchem}
\usepackage{graphicx}
\usepackage{subcaption}
\usepackage[colorinlistoftodos]{todonotes}
\usepackage[colorlinks=true, allcolors=blue]{hyperref}
\usepackage{booktabs}
\usepackage{dcolumn}
\usepackage{tabularx}
\usepackage{caption}
\usepackage[export]{adjustbox}
\usepackage{makecell}

\newcommand{\norm}[1]{\left\lVert#1\right\rVert}

\title{A universal framework for featurization of atomistic systems}

\author{
  Xiangyun Lei \\
  School of Chemical and Biomolecular Engineering\\
  Georgia Institute of Technology\\
  Atlanta, GA, 30318 USA \\
  \texttt{xlei38@gatech.edu} \\
   \And
 Andrew J. Medford\\
  School of Chemical and Biomolecular Engineering\\
  Georgia Institute of Technology\\
  Atlanta, GA, 30318 USA \\
  \texttt{ajm@gatech.edu} \\
}
\date{\today}

\begin{document}
\maketitle

\begin{abstract}
Molecular dynamics simulations are an invaluable tool in numerous scientific fields. However, the ubiquitous classical force fields cannot describe reactive systems, and quantum molecular dynamics are too computationally demanding to treat large systems or long timescales. Reactive force fields based on physics or machine learning can be used to bridge the gap in time and length scales, but these force fields require substantial effort to construct and are highly specific to a given chemical composition and application. A significant limitation of machine learning models is the use of element-specific features, leading to models that scale poorly with the number of elements. This work introduces the Gaussian multipole (GMP) featurization scheme that utilizes physically-relevant multipole expansions of the electron density around atoms to yield feature vectors that interpolate between element types and have a fixed dimension regardless of the number of elements present. We combine GMP with neural networks to directly compare it to the widely used Behler-Parinello symmetry functions for the MD17 dataset, revealing that it exhibits improved accuracy and computational efficiency. Further, we demonstrate that GMP-based models can achieve chemical accuracy for the QM9 dataset, and their accuracy remains reasonable even when extrapolating to new elements. Finally, we test GMP-based models for the Open Catalysis Project (OCP) dataset, revealing comparable performance to graph convolutional deep learning models. The results indicate that this featurization scheme fills a critical gap in the construction of efficient and transferable machine-learned force fields.
\end{abstract}

Atomistic simulations are a crucial tool in many scientific fields, ranging from protein engineering to materials design \cite{BurkePerspective,BeckeFiftyYears,MardirossianNarbe2017Tyod,HospitalAdam2015Mdsa,DeVivoMarco2016RoMD,McCammonJ.Andrew2002Mdso, HarrisonJudithA2018Roff,OganovArtemR2019Spdm,RenXinguo2012Raai}. Full quantum mechanical treatment of atoms provides highly accurate energies and forces, but the computational cost is prohibitive for the length and time scales relevant to most applications \cite{MosconiEdoardo2015AIMD,marx_hutter_2009,RaduIftimie2005AiMD}. Classical and reactive force fields can act as surrogates for the quantum mechanical simulations, enabling simulations at longer length and time scales \cite{CHARMM,AMBER,ReaxFF,OPLS,COMB, ReaxFFReview,HarrisonJudithA2018Roff, MacKerellAlexanderD2005C7EF}. However, these models are specialized to specific systems and have a limited ability to simulate inherently quantum-mechanical phenomena such as covalent bond formation \cite{ForceFieldReview}. Machine-learning models have recently emerged as a promising strategy to fill the gap between quantum mechanical simulations and classical force field models \cite{HandleyChrisM2010PESF,Behler2016PMlp,RamprasadRampi2017Mlim}. The field of machine-learned force fields has exploded in the last decade, leading to a plethora of different machine-learning force field models capable of predicting energies and forces with accuracy comparable to the underlying method \cite{BehlerParrinello,Schnet,CGCNN,DimeNet,PhysNet,HIP-NN,MXMNet,HMGNN,MPNN,SingleNN,ANI-1,FCHL,SOAP,ModifiedSOAP,wavelet}. However, most models are customized for specific application domains, and a framework for a general-purpose model has not been established.

One fundamental problem in machine-learning models for atomistic systems is the issue of feature generation. Cartesian coordinates and elemental identities are the most common way to define atomistic systems. However, this description does not capture the system's rotational, translational, or permutation invariances. Converting the Cartesian coordinates to ``feature vectors'' that describe each atom is commonly used to solve this problem. Researchers have devised a wide range of featurization strategies that encode these fundamental physical symmetries, recently classified and thoroughly covered in several thorough reviews \cite{Langer2020, Musil_2021}. Some examples include the Coulomb matrix\cite{CoulombMatrix}, radial distribution function-based fingerprint \cite{BotuVenkatesh2015Amlf}, FCHL\cite{FCHL}, SOAP \cite{SOAP, ModifiedSOAP}, COMB \cite{COMB}, Chebyshev polynomials \cite{Ceder11Species}, Gaussian momentums \cite{Gaussian_Momentum} and the ubiquitous atom-centered symmetry functions\cite{BehlerParrinello}. These strategies have yielded remarkably accurate models within various sub-fields but are not sufficiently general to treat all atomistic systems. 

The most common limitation is related to the scaling of the feature vector size as the number of elements in the system increases. Most existing descriptors scale polynomially or combinatorially with the number of elements in the system \cite{BehlerParrinello, SOAP}. This poor scaling means that these featurization strategies can only be applied to training sets with a limited number of elements, making the prospect of a universal machine-learning model that works for all elements infeasible. While some approaches overcome this limitation \cite{wavelet, Ceder11Species}, they have not been tested in on large many-element benchmark systems. In addition to fingerprint scalability, there is also the issue of scalability of the regression model. Many high-accuracy models rely on kernel ridge regression (KRR), a technique that suffers from poor scalability at with extremely large training set sizes. Neural networks are far more scalable, but if typical element-specific featurization schemes are used then a high-dimensional neural network (HDNN) \cite{BehlerParrinello,Behler_2014} is required to connect the features to atomic properties. The HDNN scheme  uses element-specific neural networks, so that the number of model parameters also increases with the number of elements. An alternative to featurization is deep learning \cite{Zhang_2018}, typically with graph-convolutional neural network (GCN) models \cite{Schnet,CGCNN,DimeNet,PhysNet,MPNN,HIP-NN,MXMNet}. These deep learning models show an excellent ability to learn appropriate representations for molecular, solid-state, and surface systems with many elements \cite{OC20}.  However, this comes at the cost of less transparent models that typically require more time for training and prediction than their feature-based counterparts. Moreover, many of the deep learning approaches utilize elemental features \cite{MPNN,Schnet,DimeNet,PhysNet}, suggesting that improved featurization schemes will translate to improved deep learning models.

Here we introduce a new featurization approach called Gaussian multipole (GMP) features. The GMP features utilize an implicit description of the electron density so that the feature vector's dimension is independent of the number of elements. Using the electron density as the fundamental input makes them suitable for universal machine-learning models that work for all elements and provides a straightforward route for extending them to systems that involve charged atoms or magnetic moments. It also naturally allows the use of the more efficient single neural network (SNN) \cite{SingleNN} structure, where all elements share the same neural network. Moreover, the GMP features are related to a multipole expansion of the electron density \cite{EdmondsA.R1957Amiq, AngularMomentum}, making them physically relevant and systematically improvable. Remarkably, we show that the GMP features are capable of interpolating between elemental species, a capability that to our knowledge has not been demonstrated with any other featurization schemes. These properties of the GMP features will facilitate a new family of fast and interpretable feature-based machine learning models that have the general applicability of more complex and opaque GCN models.

\begin{figure*}[!ht]
	\centering
    \includegraphics[width=0.98\linewidth]{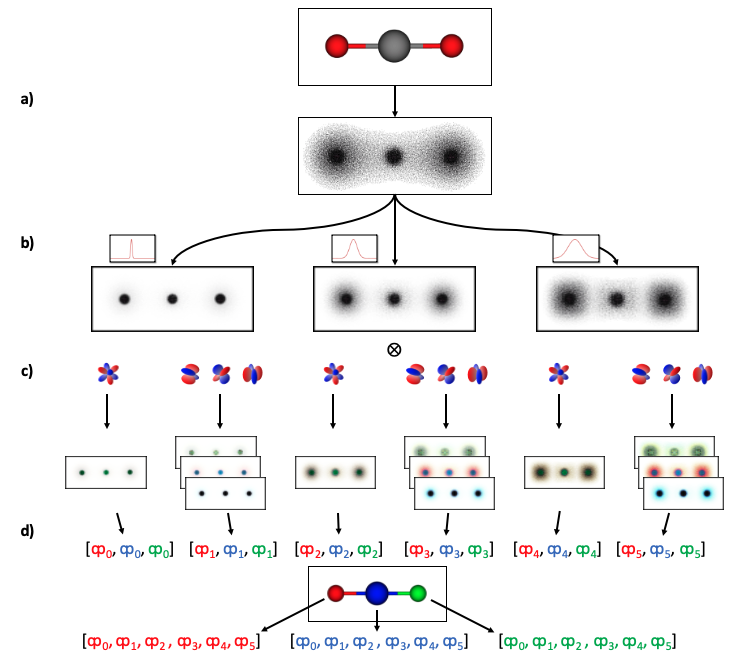}
    \caption{Illustration for the GMP featurization scheme. a) Construct the electron density distribution using linear combination of Gaussians. b) Apply the Gaussian radial probe to focus on the electronic environments at different radial length scales around each atom core by masking the electron density distribution with Gaussian functions of different widths. c) Apply the angular (MCSH) probe that acts as a multi-pole expansion of the radially-masked electron density via inner products with different groups of MCSH functions. d) Take the norm of the results for each group to ensure rotational invariance, yielding an entry to the feature vector for each individual atom.}
    \label{fig:Scheme}
\end{figure*}

The GMP approach encodes the elemental identity through an approximate description of the electron density based on Gaussian basis functions. In this work, we use the valence density extracted from the SG15 pseudopotentials \cite{SG15} approximated by 2-7 atom-centered Gaussians per element, normalized by the total number of valence electrons. The number and widths of the Gaussians for each element are determined using nonlinear regression, and these ``static valence densities'' are fixed for all models presented in this work. Details are provided in the supporting information. Conceptually, this is equivalent to using the valence density of non-interacting atoms as the fundamental description of the system (Fig. \ref{fig:Scheme}a). This approach also allows for interpolation and extrapolation between different elemental species, and concepts like charged atoms can be naturally accommodated using the isolated charged atom to construct a suitable valence potential. In principle, it is also possible to use other quantities like the all-electron density, spin density, or self-consistent electron density, as long as the quantity can be represented by a linear combination of atom-centered Gaussians. 

The approximated electron density is then vectorized via the inner product between the electron density and atom-centered ``probe'' functions to generate the features for the atoms. The probe function consists of a Gaussian function and a Maxwell-Cartesian spherical harmonic (MCSH) function. Gaussian functions of varying width account for the radial variation (Fig. \ref{fig:Scheme}b) of surrounding electron density of an atom, and the MCSH functions capture angular variation (Fig. \ref{fig:Scheme}c).  The Gaussian product rule provides analytical solutions to the integral in Eqn. \ref{eq:GMPDesc}, enabling highly efficient computation of the features. Rotational invariance is enforced by taking norms of each MCSH 
order\cite{ApplequistJon2002Mshi} (different than the previous publication \cite{MLXC2019}).
\newpage
Mathematically this is described as

\begin{equation}
\begin{split}
    \mu_{i, abc} &= < probe, \hat{\rho}> = <angular\;probe \times radial\;probe, \hat{\rho}>\\
    &= <S_{abc} \times G_{i}, \sum_{\substack{j\\atoms}} \sum_{\substack{k\\ Gaussians}} G_{dens,j,k}> \\
    &= \iiint_{V} S_{abc} G_{probe,i} \sum_{\substack{j\\atoms}} \sum_{\substack{k\\ Gaussians}} G_{dens,j,k}dV\\
    &= \sum_{\substack{j\\atoms}} \sum_{\substack{k\\ Gaussians}} \iiint_{V} S_{abc} G_{probe,i}  G_{dens,j,k}dV\\
\end{split}
\label{eq:GMPDesc}
\end{equation}

where $<a,b>$ denotes inner product of two functions, $V$ is the volume, $\mu_{i, abc}$ is the feature resulting from the radial probe $G_{i}$ and angular probe $S_{abc}$. $\hat{\rho}$ is the distribution of electron density of a molecule, approximated by linear combinations of primitive Gaussians $G_{dens,j,k}$ centered at each atom. The formalism does not include a strict cutoff radius, though for computational efficiency the sum can be restricted to surrounding atoms with a non-negligible overlap. The MCSH functions $S_{abc}$ are linear combinations of polynomials:

\begin{equation}
    S^{n}_{abc}  = \sum_{\substack{t\\terms}} C_{t} x^{m_{x,t}} y^{m_{y,t}} z^{m_{z,t}}
\label{eq:MCSH}
\end{equation}

where $abc$ is the specific index of the spherical harmonic function, $n=a+b+c$ is the order of it, and $m_x, m_y, m_z$ are the exponents of the specific polynomial. The first 4 orders of MCSH functions are listed in Table \ref{tab:MCSH}. 

\begin{table}[h]
\centering
\begin{tabular}{c c c c|c c c c}
\toprule
n & group & \{abc\} & $S^{(n)}_{abc}$ & n & group & \{abc\} & $S^{(n)}_{abc}$ \\ \midrule
0 & 1 & 000     & 1                & 3 & 1 & 300     & $15x^3-9x$\\
1 & 1 & 100     & $x$ 			   &   &   & 030     & $15y^3-9y$\\
  &   & 010     & $y$			   &   &   & 030     & $15y^3-9y$\\
  &   & 001     & $z$ 			   &   & 2 & 210     & $15x^2y-3y$\\
2 & 1 & 200     & $3x^2-1$         &   &   & 201     & $15x^2z-3z$\\
  &   & 020     & $3y^2-1$         &   &   & 021     & $15y^2z-3z$\\
  &   & 002     & $3z^2-1$         &   &   & 120     & $15xy^2-3x$\\
  & 2 & 110     & $3xy$            &   &   & 102     & $15xz^2-3x$\\
  &   & 101     & $3xz$            &   &   & 012     & $15yz^2-3y$\\
  &   & 011     & $3yz$            &   & 3 & 111     & $15xyz$\\
 \bottomrule
\end{tabular}
\caption{The analytical expressions of the first four orders of MCSH denoted by $S^{(n)}_{abc}$}
\label{tab:MCSH}
\end{table}

To ensure that the resulting features are rotationally invariant, we use the weighted sum of square of each order, $\mu_{i}$. Please note that this definition is different than our previous publication \cite{MLXC2019}. Therefore, the GMP feature vector is defined as

\begin{equation}
\vec{\textrm{\textqplig}} = \textrm{\textqplig}_{i} = \sqrt{ w_{abc} \sideset{}{_{P(a,b,c)}}\sum \mu_{i, abc}^2} \mid a,b,c \in \mathbb{N},
\label{eq:MCSHDescriptorSetDefinition}
\end{equation}

where $\textrm{\textqplig}$ denotes the GMP features, $i$ is an index over the radial probes, $abc$ is an index combination corresponding to a rotational group, and $P(a,b,c)$ denotes the permutation group of $a,b,c$ (e.g. $P(1,0,0)=\{(1,0,0), (0,1,0), (0,0,1)\}$). The weight $w_{abc}$ is shared among MCSH within the specific subgroup (perturbaion group) of order $n=a+b+c$

\begin{equation}
    w_{abc} = \frac{n!}{a!b!c!}
\end{equation}

Therefore, the set of features can thus be written as 

\begin{equation}
\vec{\textrm{\textqplig}} = \left\lbrace\begin{array}{l}
    \sqrt{\mu_{1,000}^2}, \sqrt{\mu_{1,100}^2 + \mu_{1,010}^2 + \mu_{1,001}^2}, \sqrt{(\mu_{1,200}^2 + \mu_{1,020}^2 + \mu_{1,002}^2) + 2.0( \mu_{1,110}^2 + \mu_{1,101}^2 + \mu_{1,011}^2)}, \dots \\
    
    \sqrt{\mu_{2,000}^2}, \sqrt{\mu_{2,100}^2 + \mu_{2,010}^2 + \mu_{2,001}^2}, \sqrt{(\mu_{2,200}^2 + \mu_{2,020}^2 + \mu_{2,002}^2) +2. 0(\mu_{2,110}^2 + \mu_{2,101}^2 + \mu_{2,011}^2)}, \dots \\
    \dots
  \end{array}\right\rbrace.
\label{eq:AtomisticMCSHDescriptorSet}
\end{equation}

Conceptually, the resulting features are similar to a multipole expansion of the electron density around each atom. The multipole expansion provides rotational features that are complete and orthogonal. The Gaussian probe's width controls the radial length scale of the multi-pole expansions, and the radial features are over-complete. These properties reduce linear dependencies within the features and lead to a systematic improvement in the system's description as the number of features increases. The GMP scheme is closely related to the ``wavelet'' features \cite{wavelet}, but it uses a different representation of the electron density, and differs in the way the spherical harmonics are normalized and computed, so that it provides atom-centered features. It is also similar to the SOAP scheme, which uses element-specific densities and spherical harmonics \cite{BehlerParrinello,SOAP,ModifiedSOAP}. However, the key difference is that the length of the feature vector is fixed regardless of the number elements, and the normalization across rotational groups leads to fewer features. These enable GMP to efficiently featurize complex systems with an arbitrary number of elements and arbitrary boundary conditions. 

In this work, we combine the GMP features with a neural network regression model to demonstrate the efficiency, accuracy, and transferability of models based on these features. As mentioned, the single neural network (SNN) architecture \cite{SingleNN} is well-suited to the GMP features since all elements utilize the same features (Fig. \ref{fig:Scheme}).  We also utilize a per-element bias term, equivalent to fitting to formation energies instead of total energies, to reduce the magnitude of the energies. We have implemented the GMP+SNN framework in the AMPTorch code \cite{amptorch,ShuaibiMuhammed2020Eroa} and used this implementation for all models in this work. The number of parameters used by the GMP+SNN model varies depending on the number of features, hidden layers, and nodes per layer but is generally lower than the number of parameters in a comparable GCN by an an order of magnitude. Details of model fitting, implementation, and parameters are provided in the supplementary information. The notation GMP($N_{Gaussian}$,$N_{MCSH}$) is used to denote the feature sets in the examples below, where $N_{Gaussian}$ is the number of radial Gaussians, and $N_{MCSH}$ is the maximum order MCSH used to construct the feature set. 
The notation SNN($[list\ layers]$) is used to denote the SNN model, The activation function is $Tanh$ for the MD17 test, and $GELU$ for the QM9 and OC20 tests, and batch normalization is always applied for each layer. Therefore, GMP(9,3) is a feature set constructed using 9 radial Gaussians, resulting in $9\times4=36$ features and GMP(9,3)+SNN(50,50,50) is a SNN model based on the GMP(9,3) feature set with 3 hidden layers and 50 nodes per layer. We note that models with the same number of radial Gaussian probes are not necessarily equivalent, since Gaussian widths are determined manually at this point. Additional details including the widths used for the radial Gaussians are provided in the supplementary information.

First, we compare the GMP+SNN model to the Behler-Parinello neural network (BPNN) approach that is one of the most computationally efficient featurization schemes \cite{Langer2020} and ubiquitous in materials science and chemistry due to its simplicity, generality, and efficiency \cite{BehlerParrinello,SIMPLE-NN,AMP,PhysRevB.83.153101,PhysRevB.85.045439,ArtrithNongnuch2013Nnpf, ShuaibiMuhammed2020Eroa}. We utilize an established molecular dynamics trajectory of the 3-element (C, H, O) aspirin molecule at the DFT/PBE+vdW-TS level of theory for this comparison\cite{chmiela2018}. For BPNN featurization, we use 12 variants inspired by examples in literature \cite{SchranChristoph2020AFoN}. All neural networks consist of 3 hidden layers with 50 nodes each, and the BPNN approach uses separate neural networks for each element (resulting in 3 times the number of parameters). We use a training set size of 40K images for all models, and the test and validation sets each contain 10K images. We ensure robustness by using ten randomly selected train/test/validation sets. The performance is measured by the mean absolute error (MAE) for the predicted energies of the test set images.

\begin{figure*}[!ht]
	\centering
    \begin{subfigure}{0.55\textwidth}
		\centering
        \includegraphics[width=\linewidth]{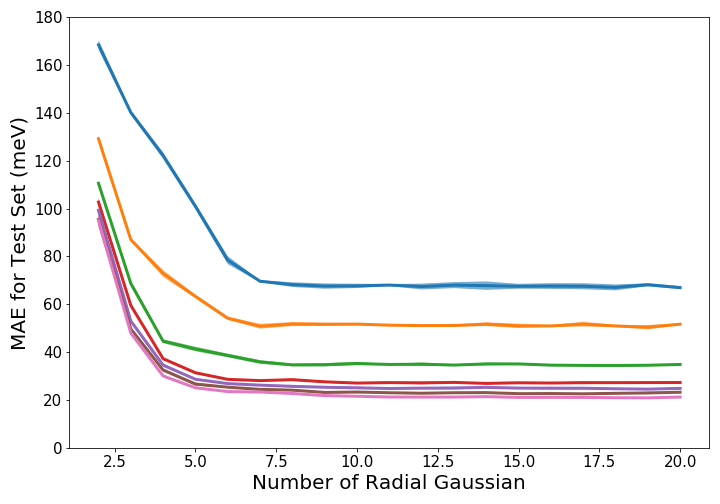}
        \caption{}
        \label{fig:MCSHSNNConvergence}
	\end{subfigure}\hspace{0.005\textwidth}
    \begin{subfigure}{0.49\textwidth}
		\centering
        \includegraphics[width=\linewidth]{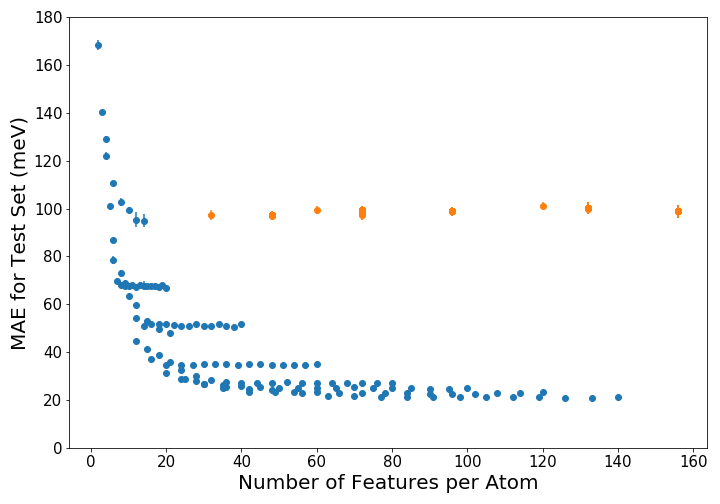}
        \caption{}
        \label{fig:maeVsNdesc}
	\end{subfigure}\hspace{0.005\textwidth}%
    \begin{subfigure}{0.49\textwidth}
		\centering
        \includegraphics[width=\linewidth]{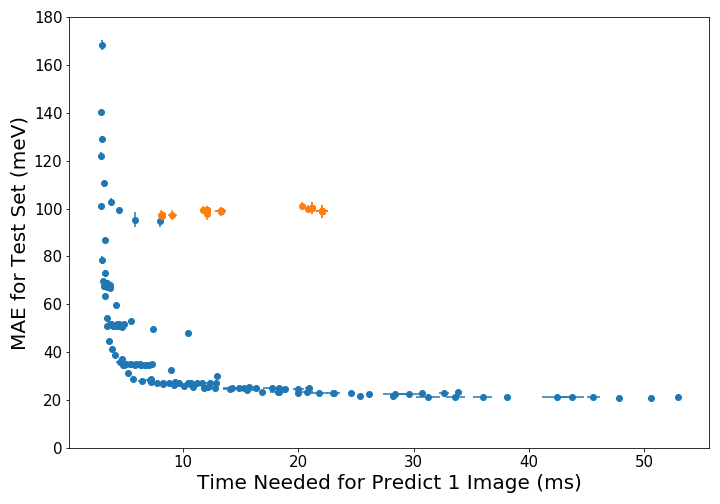}
        \caption{}
        \label{fig:maeVsUnittime}
	\end{subfigure}\hspace{0.005\textwidth}%
    \caption{Training results of GMP+SNN models compared to Behler-Parrinello + HDNN models. The training set consists of 40K images, validation set consists of 10K images and test set consists of 10K images, all randomly drawn from the MD17 aspirin simulation trajectory. Each setting was tested 10 times to ensure robustness. a) Convergence of test set MAE as rotational order and number of radial Gaussians are increased, with curves correspond to angular MCSH probes up to 1 (orange), 2 (green), 3 (red), 4 (purple), 5 (brown) and 6 (pink) orders. b) Comparison of test MAE for GMP+SNN (blue) and BPNN (orange) models as a function of the number of features. c) Comparison of test MAE for GMP+SNN and BPNN models as a function of the average prediction time for a single image.}
    \label{fig:Example1Results}
\end{figure*}

Fig \ref{fig:MCSHSNNConvergence} shows the GMP+SNN model error as a function of the multipole expansion order and the number of Gaussian radial probes. The results reveal that the GMP+SNN accuracy increases systematically with the multipole expansion order and the number of radial probes, providing a clear strategy for identifying an appropriate feature set for a given problem. In Fig \ref{fig:maeVsNdesc}, we compare the accuracy of the GMP+SNN model to the BPNN model as a function of the number of features. All Pareto-optimal models utilize the GMP+SNN approach, despite the fact that the BPNN models have three times more fitted parameters due to element-specific neural networks. It is also clear that the accuracy of BPNN models does not systematically improve with the number of features. Finally, we compare the wall time needed to compute a single image for each model. To ensure a fair comparison, we use the same CPU for each test, both approaches use the same C++ source code and loop structure, and the time is on average over the 10K predicted validation images (see supplementary information). Fig. \ref{fig:maeVsUnittime}) shows that the  GMP+SNN framework is always faster than the BPNN approach at a fixed accuracy level, or is always more accurate for a fixed computation time. This example demonstrates that the GMP+SNN approach is capable of achieving a lower error than the BPNN approach with fewer parameters and less time.

\begin{figure}[!ht]
	\centering
    \includegraphics[width=0.6\linewidth]{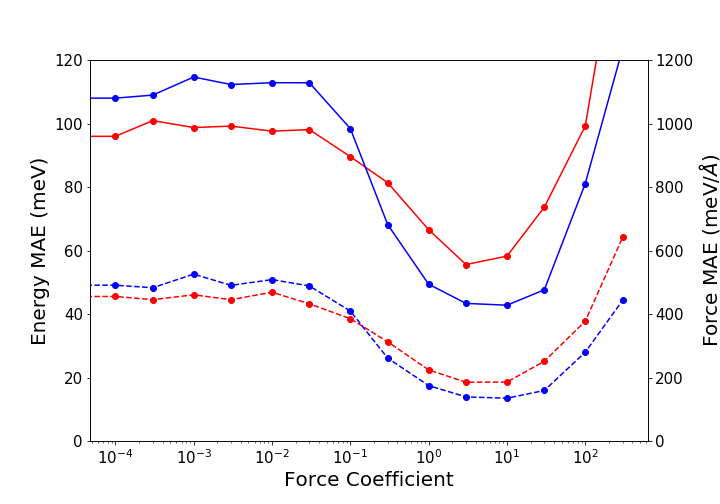}
    \caption{Test MAE of energies (solid) and forces (dashed) as a function of the force regularization coefficient on 1K randomly selected images from the aspirin trajectory. Two feature sets are compared here, both with SNN(50,3) architecture: GMP(6,3) (red, 24 features) and GMP(10,6) blue, 70 features).}
    \label{fig:forcetraining}
\end{figure}

The performance of the GMP+SNN can be further improved with force regularization, with a loss function defined as

\begin{equation}
\Gamma = \frac{1}{N_{image}} \sum_{i=1}^{N_{image}} \left\lbrace \norm{E^i_{NN}-E^i_{ref}}  + \frac{\beta}{3N^i_{atom}}\sum_{j=1}^{3N^i_{atom}} \norm{F^i_{j,NN}-F^i_{j,ref}} \right\rbrace,
\end{equation}

where $N_{image}$ is the number of training images, $N^i_{atom}$ is the number of atoms in image $i$, $E_{NN}$ and $E_{ref}$ are the predicted and reference energy of a image, $F_{NN}$ and $F_{ref}$ are the predicted and reference forces of an atom along each Cartesian axis, and $\beta$ is the force regularization coefficient. Figure \ref{fig:forcetraining} shows the results of GMP+SNN with 2 feature sets, GMP(6,3)+SNN(50,50,50) and GMP(10,6)+SNN(50,50,50), as a function of the force regularization coefficient, $\beta$. These two models are both on the Pareto frontier from the previous test. For simplicity, this test was done with 1K training images and 1K test images. The results confirm that energy training benefits from force regularization. In this case, the energy error is reduced by a factor of 2 to 3 at the optimal regularization coefficient. Both energy and force prediction accuracy show the same trend and the same optimal force coefficient, indicating that the model avoids the apparent trade-off between energy and force accuracy that has been observed for some GCN model architectures \cite{OC20}. 

\begin{figure*}[!ht]
	\centering
    \begin{subfigure}{0.48\textwidth}
		\centering
        \includegraphics[width=\linewidth]{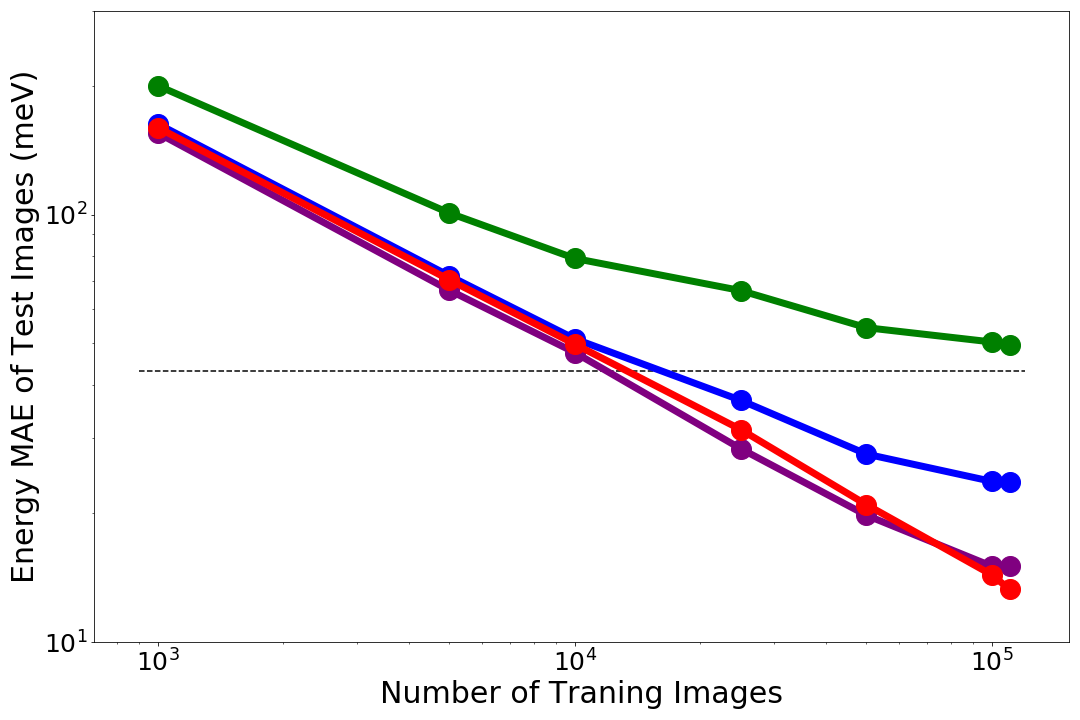}
        \caption{}
        \label{fig:QM9LC}
	\end{subfigure}\hspace{0.005\textwidth}
    \begin{subfigure}{0.48\textwidth}
		\centering
        \includegraphics[width=\linewidth]{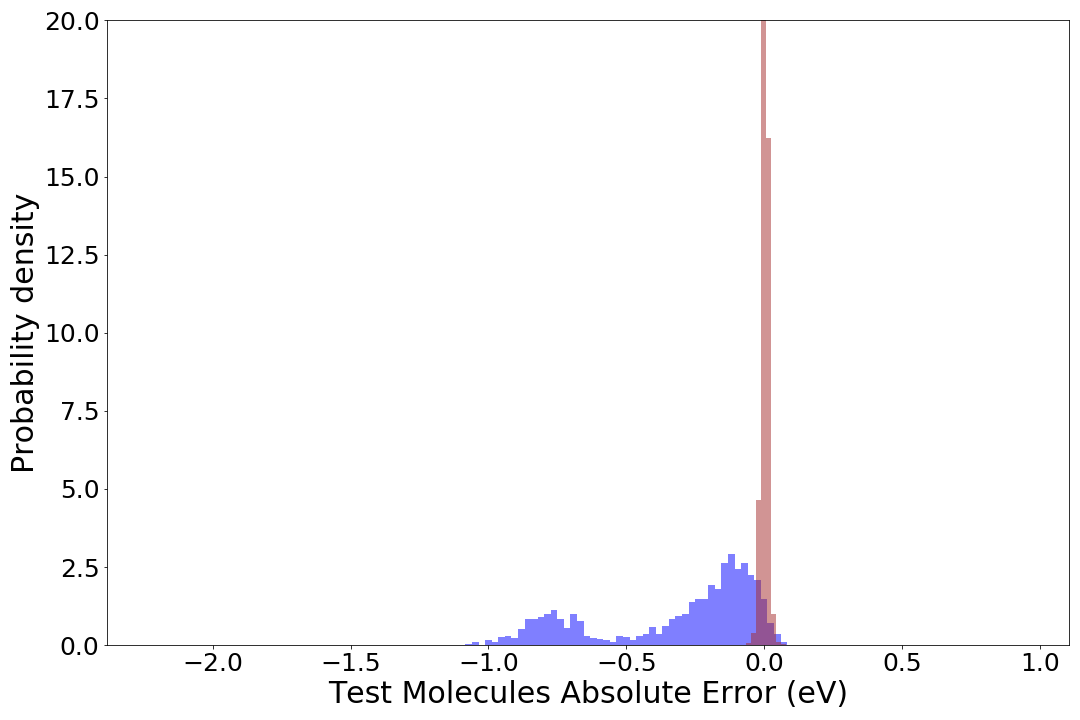}
        \caption{}
        \label{fig:QM9Transfer}
	\end{subfigure}\hspace{0.005\textwidth}%
    \caption{Learning curve and elemental transferability of GMP+SNN on the QM9 dataset. a) Learning curves of the GMP+SNN models with various feature sets and neural net sizes (green = GMP(10,1)+SNN(32,32,32), 20 features, blue = GMP(30,2)+SNN(64,64,64), 90 features, purple = GMP(50,4)+SNN(256,128,64), 250 features, red = GMP(70,6)+SNN(512,256,128,64), 490 features). The dashed line indicates chemical accuracy (43 meV). b) Test MAE distribution for GMP(70,6)+SNN(512,256,128,64) model trained exclusively on molecules without F and tested on all F-containing molecules (blue shade) and test MAE distribution for model trained and tested with randomly selected molecules (red shade).}
    \label{fig:QM9}
\end{figure*}

Next, we show that the GMP+SNN technique can scale to systems with more elements by training it on the atomization energy of the established QM9 benchmark dataset \cite{QM9}. This dataset consists of 130K chemical species with up to 9 heavy atoms and five elements (C, H, O, N, F) optimized at the B3LYP level of theory. In this example, 4 GMP descriptor sets and corresponding SNNs of different sizes are trained and tested using energies of each system. The learning curves showing the out-of-sample test error \cite{QM9Compare} as a function of training set size are shown in Fig. \ref{fig:QM9LC}. The errors of the tested GMP+SNN models are just slightly higher than the best state-of-the-art models based on Gaussian processes or GCNs (about 6 meV with 100K training data) \cite{QM9Compare,MXMNet,HMGNN,DimeNet,Schnet,PhysNet,FCHL,SOAP}, but the GMP+SNN models utilize fewer adjustable parameters and are more scalable than non-parametric models like Gaussian process regression.

Fig. \ref{fig:QM9}b shows the results of transfer learning between different elemental species in the QM9 data set. We trained a model on the set of all molecules that do not contain fluorine (128,908 molecules), and tested the same model on fluorine-containing molecules (1,923 molecules). The model error increases by around one order of magnitude when the predictions include elements not present in the training set. However, the test set MAE is 0.32 eV, which is competitive with the accuracy of generalized gradient approximation of density functional theory (GGA DFT) \cite{DFTSurvey}. This reveals that the model is reasonably accurate even for elements outside the training set. The error distribution for transfer learning is bimodal, which is primarily related to the number of F atoms in the test systems. If the error is normalized by the number of F atoms in the system the MAE decreases to 0.179 eV/F atom, and the distribution becomes more symmetric (see SI). In addition, we note that the error distribution for F atoms is relatively sensitive to the initial weights of the neural network. The MAE varies from 0.2 - 0.4 eV per system depending on initial weights
 (see SI). This indicates that the model is extrapolating and therefore more sensitive to the initial conditions and training procedure. However, the resulting models never fail catastrophically since the errors are always competitive with GGA DFT, indicating the general robustness of the transfer learning between elements.

\begin{figure*}[!ht]
	\centering
    \begin{subfigure}{0.48\textwidth}
		\centering
        \includegraphics[width=\linewidth]{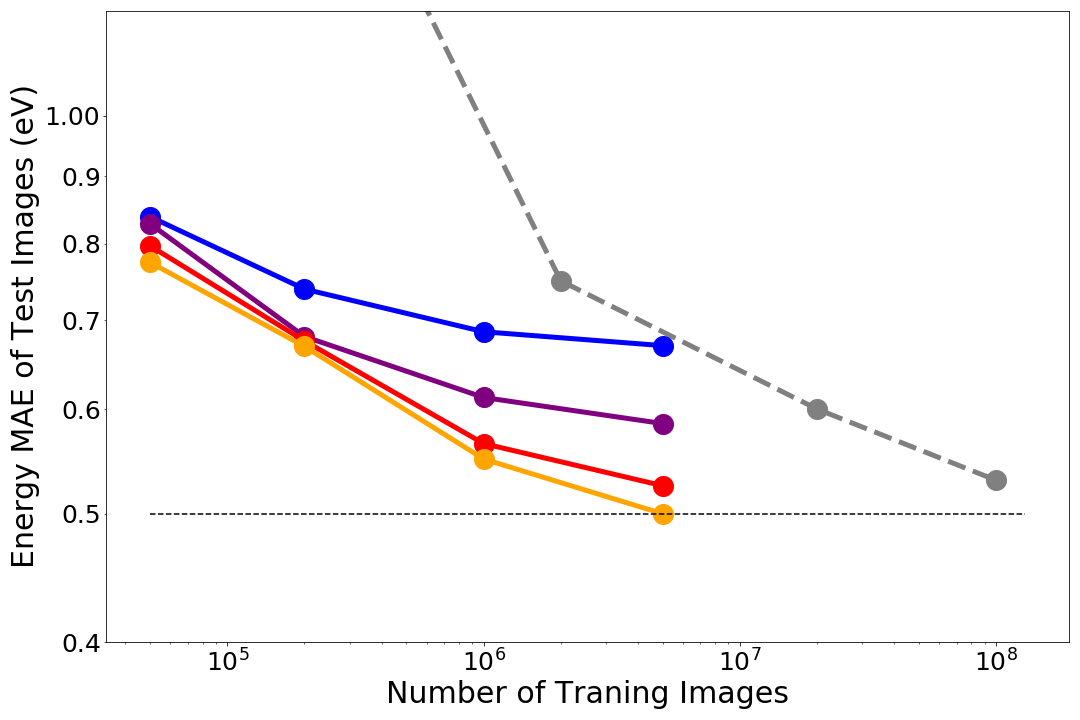}
        \caption{}
        \label{fig:OCP_id}
	\end{subfigure}\hspace{0.005\textwidth}
    \begin{subfigure}{0.48\textwidth}
		\centering
        \includegraphics[width=\linewidth]{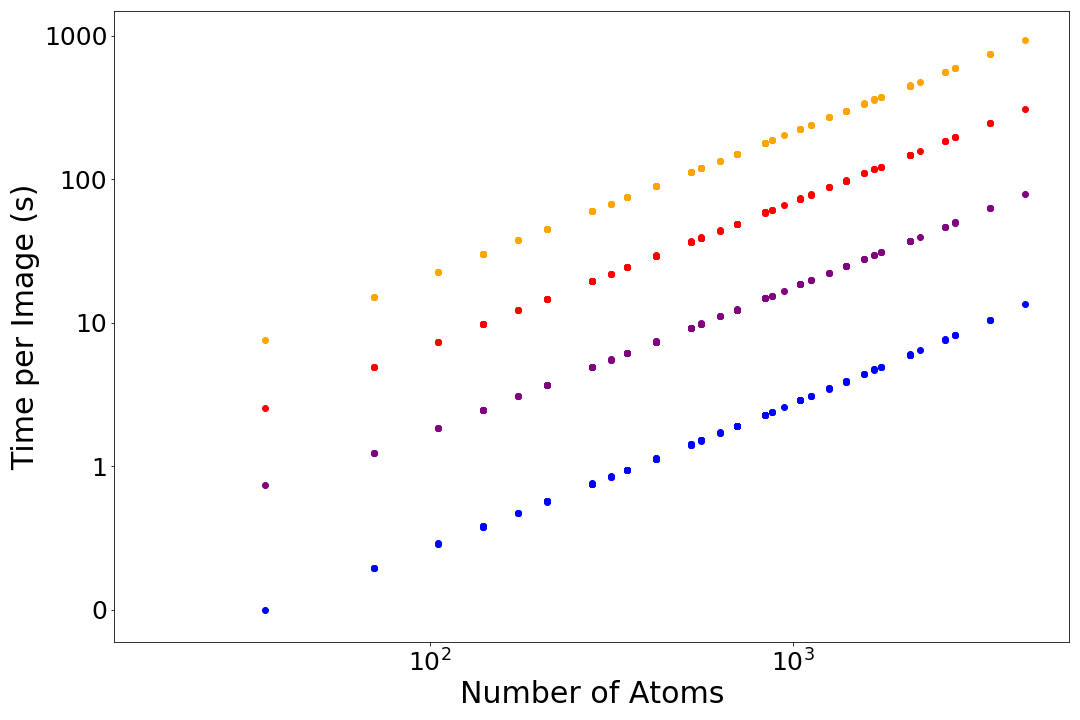}
        \caption{}
        \label{fig:OCP_scaling}
	\end{subfigure}\hspace{0.005\textwidth}
    \caption{Results for the OC20 dataset with GMP(30,2)+SNN(128,64,64) (blue, 90 features), GMP(50,4)+SNN(256,128,64) (purple, 250 features), GMP(70,6)+SNN(512,256,128,64) (red, 490 features), and GMP(90,8)+SNN(1024,512,128,64) (red, 810 features) models. a) Learning curves tested on the provided in-domain (ID) validation set  with the DimeNet++ (gray dashed) results for comparison \cite{OC20} (the gray dotted line at 0.5 eV represents an accuracy that would generally be considered useful) b) Log-log plot of time required to infer one images (with one core on a typical desktop) v.s. the number of atoms present. Systems of different sizes are made by periodically repeating one unit cell of C$_2$O adsorbed on Sn$_3$Ca (4 elements, 35 atoms).}
    \label{fig:OCP}
\end{figure*}

Finally, we test the performance of the GMP+SNN approach on the Open Catalysis Project (OC20) S2EF dataset. This dataset consists of $>$100M geometries and corresponding energies of $>$100K adsorbate-catalyst pairs containing a total of 56 elements across 82 adsorbates and up to $>100K$ different catalyst compositions for each adsorbate. The dataset also provides an independent set of 1M test systems \cite{OC20}. The energies correspond to the GGA DFT level of theory with the RPBE functional \cite{RPBE}, with mixed boundary conditions. The size, number of elements, and mix of solid-state and molecular systems make this one of the most challenging benchmark datasets available. To date, the only models capable of training and prediction for this dataset utilize elaborate GCN models \cite{OC20}. Fig. \ref{fig:OCP_id} shows the learning curves for four GMP+SNN models of different sizes tested on the provided in-domain (ID) validation set. The small model uses the GMP(30,2)+SNN(128,64,64) architecture (91 features, 34K parameters), the medium model uses the GMP(50,4)+SNN(256,128,64) architecture (250 features, 146K parameters), and the large model uses the GMP(70,6)+SNN(512,256,128,64) architecture (490 features, 528K parameters), and the largest model uses the GMP(90,8)+SNN(1024,512,128,64) architecture (810 features, 1.43M parameters). Details of all models are provided in the supplementary information. The full OC20 training set of 100M energies would require months to train with current computing resources available to us, so we restrict the analysis to energy training on data sets with fewer than 5M training images.

The results of training and testing on the OC20 set are shown in Fig. \ref{fig:OCP_id}. The in-domain test error reaches a minimum of 0.50 eV with 5M training images with the GMP(90,8)+SNN(1024,512,128,64) model, an error lower than all GCN models with 5M training points, and comparable to the performance of the CGCNN (0.527 eV) and DimeNet++ (0.486 eV) GCN models trained on all $>$100M training points \cite{OC20}. Moreover, the GMP+SNN models require fewer parameters than the GCN models, with the largest GMP+SNN model having $\sim$1.4M parameters, compared to 1.8M (DimeNet++), 3.6M (CGCNN), and 7.4M (Schnet) parameters for the GCN models\cite{OC20}. The error also decreases as the number of data points, the number of GMP features, or the neural net size increase. This suggests that further improvement is possible, although significant computational resources will be required to optimize and evaluate GMP+SNN models for the OC20 dataset.

Finally, the scaling of computational time versus the number of atoms in the system is plotted in Fig. \ref{fig:OCP_scaling}. The systems of different sizes are generated by making supercells that are periodic replications of an original unit cell of 35 atoms. The scaling is strictly linear due to the local nature of the features, which presents an advantage over most graph-based models require the entire structure as input. The linear scaling of the GMP+SNN model, along with a structure that is straightforward to parallelize, makes it ideal for the simulation of large systems that may not be feasible with graph-based models.

These examples demonstrate that the GMP featurization scheme is an efficient and universal approach to fingerprinting atomistic systems with an arbitrary number of elements or atom types. The GMP features are more computationally efficient than the deep-learning GCN models that are commonly used for many-element systems, and are even faster than the widely used Behler-Parinello symmetry functions, resulting in machine-learned force fields that can be scaled to large many-element systems and long time scales. 
The GMP feature vectors utilize an implicit description of the system's electron density, making the number of features independent of the number of elements, facilitating the inclusion of general concepts from electronic structure theory, and enabling extrapolation to new element types. Moreover, the GMP features utilize physically-meaningful concepts, leading to more interpretable models that can be systematically improved and providing a foundation for hybrid models that incorporate more physics. For example, the self-consistent electron density from a simplified Hamiltonian could be used as input to the GMP model, or the limiting behavior derived from electrostatics could be used to constrain the regression model. 

The examples presented here use the SNN regression model in conjunction with the GMP features based on static valence density to obtain accuracy comparable to state-of-the-art models on the MD17, QM9, and OC20 datasets. The results confirm that the GMP+SNN approach can reach competitive accuracy with GCN based models despite having far fewer adjustable parameters and a simpler structure.  There are numerous opportunities to revise and optimize the details of the GMP+SNN approach presented here, including modifying the valence density representation, optimizing feature selection and systematic optimization of the SNN architecture. However, the encouraging initial results across a wide range of application domains suggest that the GMP+SNN approach is a promising universal route to constructing efficient and general machine-learning models for atomistic systems.

\section*{Acknowledgments}
 This material is based upon work supported by the U.S. Department of Energy, Office of Science, Office of Basic Energy Sciences Computational Chemical Sciences program under Award Numbers DE-SC0019441 and DE-SC0019410. Computational effort was supplied partially by the National Science Foundation under Grant No. MRI-1828187. We acknowledge helpful discussions with Andrew Peterson, Zachary Ulissi, and Muhammed Shuaibi.

\typeout{}
\bibliographystyle{unsrt}
\bibliography{references}

\newpage

\section*{\begin{center}Supplementary Information for \\ ``A Universal Framework for Featurization of Atomistic Systems''\end{center}}

\subsection{Code and Saved Models}
Please checkout and install the ``MCSH\_paper1'' branch of $AMPTorch$ to try any of the saved models and test scripts in this study. The code can be found here: \url{ https://github.com/medford-group/amptorch}. The test scripts can be found here: \url{https://github.com/medford-group/GMP_AmpTorch_Tests}. Tutorials for regenerating all the test models are all included in the repo.

\newpage

\subsection{Sample Calculation and Proof of Rotation Invariance}
To show that the norms of the groups are rotation invariance, it is suffice to show that the results depend only on distances.

\subsubsection{General Expression}

Substitute in the definition of $S_{abc}$ and the Gaussian functions:
\begin{equation}
\begin{split}
    \mu_{i, abc} &= < probe, \hat{\rho}> = <angular\;probe \times radial\;probe, \hat{\rho}>\\
    &= \sum_{\substack{j\\atoms}} \sum_{\substack{k\\ Gaussians}} \iiint_{V} S_{abc} G_{probe,i}  G_{dens,j,k}dV\\
    &= \sum_{\substack{j\\atoms}} \sum_{\substack{k\\ Gaussians}} \iiint_{V} \sum_{\substack{t\\terms}} C_{t} x^{m_{x,t}} y^{m_{y,t}} z^{m_{z,t}} A_ie^{-\alpha_i r^2}  B_{j,k}e^{-\beta_{j,k}(r-r_{0;j})^2} dV\\
\end{split}
\label{eq:GMPcalc1}
\end{equation}

where $A, B, \alpha, \beta$ are parameters for the Gaussian functions, and $r_0$'s are the relative coordinates of the nearby atoms of interest.

Since the product of Gaussian functions is another Gaussian function, the equation can be written as:

\begin{equation}
\begin{split}
    \mu_{i, abc} 
    &= \sum_{j} \sum_{k} \sum_{t} \iiint_{V} C_{t} x^{m_{x,t}} y^{m_{y,t}} z^{m_{z,t}}   D_{i,j,k}e^{-\gamma_{i,j,k}(r-r'_{0;i,j,k})^2} dV\\
    &= \sum_{j,k,t} K_{i,j,k,t} 
    \int_{-\infty}^{\infty} x^{m_{x,t}} e^{-\gamma_{i,j,k}(x-x'_{0;i,j,k})^2} dx 
    \int_{-\infty}^{\infty} y^{m_{y,t}} e^{-\gamma_{i,j,k}(y-y'_{0;i,j,k})^2} dy 
    \int_{-\infty}^{\infty} z^{m_{z,t}} e^{-\gamma_{i,j,k}(z-z'_{0;i,j,k})^2} dz
\end{split}
\label{eq:GMPcalc2}
\end{equation}

where $D,K$ are constants, $r'_0$'s are the center of the resulting Gaussian functions, and $x'_0, y'_0,z'_0$ are the components of $r'_0$. Since the integral
\begin{equation}
    \int_{-\infty}^{\infty} x^{m} e^{-\gamma(x-x'_{0})^2} dx 
\label{eq:GMPcalc3}
\end{equation}
has analytical solutions, $\mu_{i, abc}$ is straightforward to calculate. More specifically, 

\begin{equation}
\begin{split}
    \int_{-\infty}^{\infty} x^{0} e^{-\gamma(x-x'_{0})^2} dx & = \sqrt{\frac{\pi}{\gamma}}\\
    \int_{-\infty}^{\infty} x^{1} e^{-\gamma(x-x'_{0})^2} dx & = x'_{0}\sqrt{\frac{\pi}{\gamma}}\\
    \int_{-\infty}^{\infty} x^{2} e^{-\gamma(x-x'_{0})^2} dx & = (\frac{1}{2\gamma} + x_0^{'2})\sqrt{\frac{\pi}{\gamma}}\\
    ...
\end{split}
\label{eq:GMPcalc4}
\end{equation}

\subsection{Rotation Invariance}
To show the features $\textrm{\textqplig}_{i,abc}$ are rotation invariance, it is suffice to show that the values of them only depend on distances. Although a general proof that it works for all MCSH functions ($S^n_{abc}$) is more involved and beyond the scope of this work, it is straight forward to show few specific cases as examples, and the rest can be proved the same way.

\subsubsection{$\textrm{\textqplig}_{i,000}$}
Follow Equation \ref{eq:GMPcalc2}, and note that $S_{000}=1$ only has one term
\begin{equation}
\begin{split}
    \textrm{\textqplig}_{i,000} &= \mu_{i, 000} \\
    &= \sum_{j,k} K_{i,j,k} 
    \int_{-\infty}^{\infty} e^{-\gamma_{i,j,k}(x-x'_{0;i,j,k})^2} dx 
    \int_{-\infty}^{\infty} e^{-\gamma_{i,j,k}(y-y'_{0;i,j,k})^2} dy 
    \int_{-\infty}^{\infty} e^{-\gamma_{i,j,k}(z-z'_{0;i,j,k})^2} dz\\
    &= \sum_{j,k} K_{i,j,k} (\frac{\pi}{\gamma_{i,j,k}})^{\frac{3}{2}}
\end{split}
\label{eq:GMPcalcRI1}
\end{equation}
It is evident that this feature is rotation invariant

\subsubsection{$\textrm{\textqplig}_{i,100}$}
Follow Equation \ref{eq:GMPcalc2}
\begin{equation}
\begin{split}
    \textrm{\textqplig}_{i,100} &= \sqrt{\mu_{i, 100}^2 + \mu_{i, 010}^2 + \mu_{i, 001}^2}\\
    S_{100}&=x\\
    S_{010}&=y\\
    S_{001}&=z\\
\end{split}
\label{eq:GMPcalcRI2}
\end{equation}

where
\begin{equation}
\begin{split}
    \mu_{i, 100} &= \sum_{j,k} K_{i,j,k} 
    \int_{-\infty}^{\infty} x e^{-\gamma_{i,j,k}(x-x'_{0;i,j,k})^2} dx 
    \int_{-\infty}^{\infty} e^{-\gamma_{i,j,k}(y-y'_{0;i,j,k})^2} dy 
    \int_{-\infty}^{\infty} e^{-\gamma_{i,j,k}(z-z'_{0;i,j,k})^2} dz\\
    &= \sum_{j,k} K_{i,j,k} x'_{0;i,j,k} (\frac{\pi}{\gamma_{i,j,k}})^{\frac{3}{2}}\\
    &= \sum_{l=1} K_{i,l} x'_{0;i,l} (\frac{\pi}{\gamma_{i,l}})^{\frac{3}{2}}\\
    \mu_{i, 100}^2 &= K_{i,1}^2 x^{'2}_{0;i,1} + K_{i,1}K_{i,2} x'_{0;i,l}x'_{0;i,2} + ...
\end{split}
\label{eq:GMPcalcRI3}
\end{equation}

where $l$ iterates through $j,k$. similarly,
\begin{equation}
\begin{split}
    \mu_{i, 010}^2 &= K_{i,1}^2 y^{'2}_{0;i,1} + K_{i,1}K_{i,2} y'_{0;i,l}y'_{0;i,2} + ...\\
    \mu_{i, 001}^2 &= K_{i,1}^2 z^{'2}_{0;i,1} + K_{i,1}K_{i,2} z'_{0;i,l}z'_{0;i,2} + ...\\
\end{split}
\label{eq:GMPcalcRI4}
\end{equation}

Threfore,
\begin{equation}
\begin{split}
    \textrm{\textqplig}_{i,100} &= \sqrt{\mu_{i, 100}^2 + \mu_{i, 010}^2 + \mu_{i, 001}^2} \\
    &= \sqrt{K_{i,1}^2(x^{'2}_{0;i,1} + y^{'2}_{0;i,1} + z^{'2}_{0;i,1}) 
    + K_{i,1}K_{i,2} (x'_{0;i,l}x'_{0;i,2} + y'_{0;i,l}y'_{0;i,2} + z'_{0;i,l}z'_{0;i,2}) + ...} \\
    &= \sqrt{K_{i,1}^2 <r'_{0;i,l}, r'_{0;i,l}> 
    + K_{i,1}K_{i,2} <r'_{0;i,l}, r'_{0;i,2}> + ...}
\end{split}
\label{eq:GMPcalcRI4}
\end{equation}
which is only a function of distances. Hence it is rotation invariant.

\newpage
\subsection{MD17 Aspirin Examples}

\subsubsection{Standard Training Procedure}
Cutoffs for both the BP scheme and the GMP scheme are set to 10 \AA{}. The neural networks are all trained using the same procedure: $lr = 1e^{-3}$ for $6000$ epochs. For the BP vs. GMP comparison example on aspirin MD data, the batch size is chosen to be 256 images. For the force training example, the batch size is chosen to be 32 images.

\subsubsection{Behler-Parrinello + HDNN Comparison Test Setups}
12 sets of Behler-Parrinello feature are selected, as listed below

\begin{table*}[!h]
\centering
\begin{tabular}{c | c c | c c c | c} 
\toprule
Set & \multicolumn{2}{c|}{G2} & \multicolumn{3}{c|}{G4} & $N_{feature}$$^a$\\
\midrule
  & $\eta$ & $R_s$ & $\eta$ & $\zeta$ & $\gamma$ \\
\midrule

1 & \small\makecell{[0.05, 0.0965, 0.1864, 0.3598, \\ 0.6947, 1.3413 , 2.5897, 5.0]} & \small[0, 1.5] & \small[0.001, 0.01, 0.03] & \small[1.0, 2.0, 4.0] & \small[1.0, -1.0] & 156 \\ \midrule

2 & \small\makecell{[0.05, 0.0965, 0.1864, 0.3598, \\ 0.6947, 1.3413 , 2.5897, 5.0]} & \small[0, 1.5] & \small[0.01, 0.03] & \small[1.0,  4.0] & \small[1.0, -1.0] & 96 \\ \midrule

3 & \small\makecell{[0.05, 0.0965, 0.1864, 0.3598, \\ 0.6947, 1.3413 , 2.5897, 5.0]} & \small[0] & \small[0.01] & \small[1.0, 4.0] & \small[1.0, -1.0] & 48 \\ \midrule

4 & \small\makecell{[0.05, 0.0965, 0.1864, 0.3598, \\ 0.6947, 1.3413 , 2.5897, 5.0]} & \small[0] & \small[0.001, 0.01, 0.03] & \small[1.0, 2.0, 4.0] & \small[1.0, -1.0] & 132 \\ \midrule

5 & \small\makecell{[0.05, 0.0965, 0.1864, 0.3598, \\ 0.6947, 1.3413 , 2.5897, 5.0]} & \small[0] & \small[ 0.01, 0.03] & \small[1.0, 4.0] & \small[1.0, -1.0] & 72 \\ \midrule

6 & \small\makecell{[0.05, 0.0965, 0.1864, 0.3598, \\ 0.6947, 1.3413 , 2.5897, 5.0]} & \small[0, 1.5] & \small[0.01] & \small[1.0, 4.0] & \small[1.0, -1.0] & 72 \\ \midrule

7 & \small[0.05, 0.2324, 1.0772, 5.] & \small[0, 1.5] & \small[0.001, 0.01, 0.03] & \small[1.0, 2.0, 4.0] & \small[1.0, -1.0] & 132 \\ \midrule

8 & \small[0.05, 0.2324, 1.0772, 5.] & \small[0, 1.5] & \small[0.01, 0.03] & \small[1.0, 4.0] & \small[1.0, -1.0] & 72 \\ \midrule

9 & \small[0.05, 0.2324, 1.0772, 5.] & \small[0] & \small[ 0.01] & \small[1.0, 4.0] & \small[1.0, -1.0] & 32 \\ \midrule

10 & \small[0.05, 0.2324, 1.0772, 5.] & \small[0] & \small[0.001, 0.01, 0.03] & \small[1.0, 2.0, 4.0] & \small[1.0, -1.0] & 120 \\ \midrule

11 & \small[0.05, 0.2324, 1.0772, 5.] & \small[0] & \small[0.01, 0.03] & \small[1.0, 4.0] & \small[1.0, -1.0] & 60 \\ \midrule

12 & \small[0.05, 0.2324, 1.0772, 5.] & \small[0, 1.5] & \small[0.01] & \small[1.0, 4.0] & \small[1.0, -1.0] & 48 \\ 
 \bottomrule
\end{tabular}
\caption{List of tested Behler-Parrinello feature sets. $\eta$ and $R_s$ for G2 functions are used combinatorially, same as $\eta$, $\zeta$ and $\gamma$ for G4 functions. Moreover, there are 3 types of elements (C, H, O) for this dataset. Therefore, feature set 1 has $3 (elements)\times8 (\eta) \times 2 (R_s) + 6(\,possible\, element\, pairs) \times 3 (\eta) \times 3 (\zeta)s \times 2 (\gamma) = 156$ features per atom. $^a$Number of features per atom.}
\label{tab:listMCSHGroups}
\end{table*}

The list of test results with BP + HDNN models are shown below in Table \ref{tab:BPHDNNResults}

\begin{table*}[]
\centering
\begin{tabular}{c c c c c}
\toprule
Set & $N_{feature}$$^a$ & MAE train (meV) & MAE test (meV) & Time (ms/image) \\ \midrule
1	&	156	&	$73.0\pm3.0$	&	$98.8\pm2.7$	&	$22.1\pm0.5$ \\
2	&	96	&	$74.2\pm1.4$	&	$98.8\pm1.7$	&	$13.3\pm0.4$ \\
3	&	48	&	$71.8\pm1.9$	&	$97.3\pm1.8$	&	$8.2\pm0.1$ \\
4	&	132	&	$73.5\pm1.6$	&	$100.3\pm2.5$	&	$21.2\pm0.3$ \\
5	&	72	&	$73.2\pm1.2$	&	$99.6\pm1.5$	&	$12.1\pm0.1$ \\
6	&	72	&	$74.4\pm1.6$	&	$97.4\pm2.2$	&	$9.1\pm0.4$ \\
7	&	132	&	$73.3\pm1.4$	&	$99.9\pm1.5$	&	$20.9\pm0.1$ \\
8	&	72	&	$71.6\pm1.5$	&	$97.8\pm2.4$	&	$12.1\pm0.1$ \\
9	&	32	&	$70.5\pm2.2$	&	$97.3\pm1.9$	&	$8.1\pm0.2$ \\
10	&	120	&	$73.5\pm1.5$	&	$101.2\pm1.4$	&	$20.4\pm0.0$ \\
11	&	60	&	$71.7\pm2.1$	&	$99.3\pm1.6$	&	$11.8\pm0.1$ \\
12	&	48	&	$72.4\pm1.7$	&	$96.9\pm1.3$	&	$8.2\pm0.1$ \\

 \bottomrule
\end{tabular}
\caption{Performance test results of the tested GMP + HDNN setups. The values are the average values of the 10 trials, and the uncertainties are estimated by their standard deviation. $^a$Number of features per atom.}
\label{tab:BPHDNNResults}
\end{table*}

\newpage
\subsubsection{GMP+SNN Comparison Test Setups}
The probe of GMP has two parts: groups of MCSHs for probing angular features and radial gaussian for probing radial features. In this work, all orders of MCSH up to the indicated order are included (starting from order 0), and the number of possible groups for each order are listed below in Table \ref{tab:listMCSHGroups2}:

We combine the radial probes with the lists of radial Gaussians combinatorially to obtain the full list of probes/features. The list of widths (standard deviations) of the probe Gaussians is simply chosen to be uniform spaced values up to 2.0 (in the language of python, it is: linspace(0, 2.0, n\_gaussian+1, endpoint=True)[1:])

Therefore, when there are 5 Gaussians with MCSH up to order 6, there are $5\times7=35$ descriptors per atom. The complete list of test results with GMP+SNN is give in Table \ref{tab:GMPSNNResultFull}.


\begin{table*}[]
\centering
\resizebox{0.5\textwidth}{!}{%
\begin{tabular}{c c c c c c}
\toprule
Num. Gaussians$^a$ & MCSH order$^b$ & $N_{feature}$$^c$ & MAE train (meV) & MAE test (meV) & Time (ms/image) \\ \midrule
2	&	0	&	2	&	$164.9\pm0.7$	&	$168.3\pm2.0$	&	$2.9\pm0.3$ \\
2	&	1	&	4	&	$117.3\pm1.1$	&	$129.1\pm1.3$	&	$2.9\pm0.1$ \\
2	&	2	&	6	&	$94.8\pm1.2$	&	$110.5\pm0.8$	&	$3.2\pm0.2$ \\
2	&	3	&	8	&	$84.2\pm0.8$	&	$102.7\pm1.7$	&	$3.8\pm0.2$ \\
2	&	4	&	10	&	$79.2\pm1.0$	&	$99.3\pm1.3$	&	$4.5\pm0.0$ \\
2	&	5	&	12	&	$76.0\pm2.3$	&	$95.4\pm3.2$	&	$5.8\pm0.0$ \\
2	&	6	&	14	&	$74.2\pm1.7$	&	$95.0\pm2.6$	&	$8.0\pm0.3$ \\
2	&	7	&	16	&	$71.3\pm1.0$	&	$91.4\pm1.3$	&	$11.2\pm0.3$ \\
2	&	8	&	18	&	$72.5\pm0.7$	&	$93.7\pm0.6$	&	$14.9\pm0.3$ \\ \midrule
4	&	0	&	4	&	$117.7\pm0.6$	&	$122.1\pm1.4$	&	$2.9\pm0.1$ \\
4	&	1	&	8	&	$68.2\pm1.5$	&	$72.9\pm1.4$	&	$3.2\pm0.1$ \\
4	&	2	&	12	&	$40.6\pm0.6$	&	$44.6\pm0.8$	&	$3.6\pm0.0$ \\
4	&	3	&	16	&	$33.2\pm0.6$	&	$37.3\pm0.5$	&	$4.7\pm0.1$ \\
4	&	4	&	20	&	$30.3\pm0.7$	&	$34.7\pm0.7$	&	$6.4\pm0.1$ \\
4	&	5	&	24	&	$28.2\pm0.4$	&	$32.6\pm0.6$	&	$8.9\pm0.0$ \\
4	&	6	&	28	&	$26.1\pm0.5$	&	$30.1\pm0.5$	&	$12.9\pm0.2$ \\
4	&	7	&	32	&	$24.3\pm0.3$	&	$28.5\pm0.4$	&	$19.3\pm0.0$ \\
4	&	8	&	36	&	$23.5\pm0.2$	&	$27.5\pm0.5$	&	$27.2\pm0.2$ \\ \midrule
6	&	0	&	6	&	$76.3\pm1.6$	&	$78.5\pm1.7$	&	$3.0\pm0.2$ \\
6	&	1	&	12	&	$50.4\pm0.4$	&	$54.2\pm0.7$	&	$3.4\pm0.2$ \\
6	&	2	&	18	&	$35.2\pm0.7$	&	$38.7\pm0.7$	&	$4.1\pm0.1$ \\
6	&	3	&	24	&	$25.5\pm0.3$	&	$28.7\pm0.2$	&	$5.7\pm0.1$ \\
6	&	4	&	30	&	$24.1\pm0.4$	&	$26.9\pm0.4$	&	$8.3\pm0.2$ \\
6	&	5	&	36	&	$22.1\pm0.4$	&	$25.5\pm0.3$	&	$12.2\pm0.3$ \\
6	&	6	&	42	&	$20.4\pm0.2$	&	$23.6\pm0.1$	&	$18.2\pm0.7$ \\
6	&	7	&	48	&	$19.8\pm0.6$	&	$23.1\pm0.4$	&	$27.8\pm0.3$ \\
6	&	8	&	54	&	$19.1\pm0.4$	&	$22.5\pm0.5$	&	$39.1\pm0.2$ \\ \midrule
8	&	0	&	8	&	$66.9\pm0.9$	&	$68.2\pm1.0$	&	$3.3\pm0.2$ \\
8	&	1	&	16	&	$48.3\pm1.0$	&	$51.8\pm0.8$	&	$3.7\pm0.3$ \\
8	&	2	&	24	&	$32.0\pm0.2$	&	$34.8\pm0.5$	&	$4.9\pm0.4$ \\
8	&	3	&	32	&	$25.7\pm0.3$	&	$28.6\pm0.4$	&	$7.1\pm0.4$ \\
8	&	4	&	40	&	$23.0\pm0.3$	&	$25.8\pm0.4$	&	$10.1\pm0.4$ \\
8	&	5	&	48	&	$21.0\pm0.5$	&	$24.2\pm0.4$	&	$15.5\pm0.6$ \\
8	&	6	&	56	&	$20.0\pm0.3$	&	$22.8\pm0.1$	&	$23.1\pm0.6$ \\
8	&	7	&	64	&	$19.3\pm0.2$	&	$22.3\pm0.2$	&	$36.1\pm1.0$ \\
8	&	8	&	72	&	$18.0\pm0.4$	&	$21.0\pm0.4$	&	$51.1\pm0.8$ \\ \midrule
10	&	0	&	10	&	$66.3\pm1.0$	&	$67.7\pm0.9$	&	$3.2\pm0.1$ \\
10	&	1	&	20	&	$48.3\pm0.2$	&	$51.8\pm0.4$	&	$3.7\pm0.0$ \\
10	&	2	&	30	&	$32.4\pm0.6$	&	$35.3\pm0.7$	&	$5.0\pm0.2$ \\
10	&	3	&	40	&	$24.4\pm0.4$	&	$27.2\pm0.4$	&	$7.8\pm0.1$ \\
10	&	4	&	50	&	$22.4\pm0.5$	&	$25.2\pm0.5$	&	$11.8\pm0.1$ \\
10	&	5	&	60	&	$20.6\pm0.3$	&	$23.4\pm0.3$	&	$18.4\pm0.1$ \\
10	&	6	&	70	&	$18.8\pm0.2$	&	$21.7\pm0.3$	&	$28.2\pm0.3$ \\
10	&	7	&	80	&	$18.1\pm0.3$	&	$20.9\pm0.3$	&	$44.0\pm0.6$ \\
10	&	8	&	90	&	$16.8\pm0.2$	&	$19.7\pm0.2$	&	$63.9\pm1.3$ \\ \midrule
12	&	0	&	12	&	$65.9\pm1.0$	&	$67.5\pm1.2$	&	$3.4\pm0.2$ \\
12	&	1	&	24	&	$47.8\pm0.7$	&	$51.2\pm0.4$	&	$3.9\pm0.0$ \\
12	&	2	&	36	&	$32.3\pm1.1$	&	$35.0\pm0.7$	&	$5.5\pm0.2$ \\
12	&	3	&	48	&	$24.4\pm0.6$	&	$27.2\pm0.6$	&	$8.8\pm0.2$ \\
12	&	4	&	60	&	$22.2\pm0.3$	&	$25.1\pm0.4$	&	$14.2\pm0.8$ \\
12	&	5	&	72	&	$20.3\pm0.3$	&	$22.9\pm0.4$	&	$21.8\pm0.8$ \\
12	&	6	&	84	&	$18.6\pm0.4$	&	$21.3\pm0.3$	&	$33.6\pm0.8$ \\
12	&	7	&	96	&	$17.7\pm0.3$	&	$20.5\pm0.3$	&	$52.0\pm0.6$ \\
12	&	8	&	108	&	$17.0\pm0.3$	&	$19.7\pm0.1$	&	$74.9\pm1.9$ \\ \midrule
14	&	0	&	14	&	$66.2\pm1.6$	&	$67.8\pm1.8$	&	$3.6\pm0.3$ \\
14	&	1	&	28	&	$48.7\pm0.7$	&	$51.7\pm0.8$	&	$4.2\pm0.1$ \\
14	&	2	&	42	&	$32.3\pm0.5$	&	$35.2\pm0.6$	&	$6.0\pm0.1$ \\
14	&	3	&	56	&	$24.2\pm0.3$	&	$27.0\pm0.3$	&	$9.7\pm0.1$ \\
14	&	4	&	70	&	$22.6\pm0.2$	&	$25.4\pm0.3$	&	$15.7\pm0.7$ \\
14	&	5	&	84	&	$20.3\pm0.2$	&	$23.2\pm0.3$	&	$24.5\pm0.1$ \\
14	&	6	&	98	&	$18.8\pm0.1$	&	$21.5\pm0.1$	&	$38.1\pm0.3$ \\
14	&	7	&	112	&	$17.8\pm0.3$	&	$20.7\pm0.2$	&	$60.1\pm0.7$ \\
14	&	8	&	126	&	$16.8\pm0.3$	&	$19.6\pm0.2$	&	$88.6\pm3.5$ \\ \midrule
16	&	0	&	16	&	$66.5\pm1.3$	&	$67.7\pm1.2$	&	$3.5\pm0.1$ \\
16	&	1	&	32	&	$47.7\pm0.4$	&	$51.0\pm0.4$	&	$4.3\pm0.0$ \\
16	&	2	&	48	&	$32.1\pm0.3$	&	$34.7\pm0.4$	&	$6.4\pm0.1$ \\
16	&	3	&	64	&	$24.5\pm0.3$	&	$27.2\pm0.1$	&	$10.7\pm0.1$ \\
16	&	4	&	80	&	$22.3\pm0.5$	&	$25.0\pm0.6$	&	$17.7\pm0.8$ \\
16	&	5	&	96	&	$20.0\pm0.3$	&	$22.8\pm0.3$	&	$28.4\pm1.0$ \\
16	&	6	&	112	&	$18.6\pm0.1$	&	$21.2\pm0.2$	&	$43.7\pm1.1$ \\
16	&	7	&	128	&	$17.9\pm0.2$	&	$20.6\pm0.4$	&	$69.5\pm2.0$ \\
16	&	8	&	144	&	$16.9\pm0.2$	&	$19.7\pm0.3$	&	$101.5\pm5.0$ \\ \midrule
18	&	0	&	18	&	$65.6\pm1.2$	&	$67.1\pm1.2$	&	$3.6\pm0.0$ \\
18	&	1	&	36	&	$47.7\pm0.4$	&	$51.0\pm0.4$	&	$4.7\pm0.1$ \\
18	&	2	&	54	&	$32.0\pm0.4$	&	$34.5\pm0.6$	&	$6.8\pm0.0$ \\
18	&	3	&	72	&	$24.6\pm0.3$	&	$27.4\pm0.2$	&	$11.7\pm0.1$ \\
18	&	4	&	90	&	$22.0\pm0.2$	&	$24.8\pm0.3$	&	$18.9\pm0.0$ \\
18	&	5	&	108	&	$20.2\pm0.2$	&	$22.9\pm0.3$	&	$30.7\pm0.1$ \\
18	&	6	&	126	&	$18.4\pm0.3$	&	$21.1\pm0.2$	&	$47.9\pm0.0$ \\
18	&	7	&	144	&	$17.5\pm0.4$	&	$20.3\pm0.3$	&	$75.8\pm0.2$ \\
18	&	8	&	162	&	$16.7\pm0.1$	&	$19.5\pm0.2$	&	$109.2\pm0.1$ \\ \midrule
20	&	0	&	20	&	$65.3\pm0.4$	&	$67.0\pm0.6$	&	$3.7\pm0.0$ \\
20	&	1	&	40	&	$48.7\pm0.5$	&	$51.7\pm0.4$	&	$4.9\pm0.2$ \\
20	&	2	&	60	&	$31.9\pm0.3$	&	$34.9\pm0.4$	&	$7.3\pm0.1$ \\
20	&	3	&	80	&	$24.7\pm0.3$	&	$27.4\pm0.1$	&	$12.8\pm0.4$ \\
20	&	4	&	100	&	$22.2\pm0.5$	&	$25.0\pm0.5$	&	$20.9\pm0.3$ \\
20	&	5	&	120	&	$20.4\pm0.2$	&	$23.3\pm0.2$	&	$33.9\pm0.1$ \\
20	&	6	&	140	&	$18.6\pm0.4$	&	$21.3\pm0.4$	&	$53.0\pm0.1$ \\
20	&	7	&	160	&	$17.8\pm0.4$	&	$20.6\pm0.4$	&	$83.7\pm0.1$ \\
20	&	8	&	180	&	$16.9\pm0.2$	&	$19.6\pm0.2$	&	$121.3\pm1.0$ \\
 \bottomrule
\end{tabular}}
\caption{Performance test results of the full sets of tested GMP + SNN setups. The values are the average values of the 10 trials, and the uncertainties are estimated by their standard deviation. $^a$Number of possible Gaussian functions used to construct the descriptor probes. $^b$The highest MCSH order used to construct the probes. For example, when highest order is 2, that means all groups from MCSH of order 0, 1 and 2 are used to construct the probes. $^c$Number of features per atom.}
\label{tab:GMPSNNResultFull}
\end{table*}

\newpage

\subsubsection{Force Training Example}
The sigmas of the radial probe Gaussians are listed in Table \ref{tab:forcetrainingSetup}
\begin{table*}
\centering
\begin{tabular}{c c c}
\toprule
Model & Sigmas & $N_{feature}$ \\ \midrule
$GMP(6,3)+SNN(50,3)$ & [0.333, 0.666, 1.0, 1.333, 1.666, 2.0] & 24 \\ \midrule
$GMP(10,6)+SNN(50,3)$ & [0.2,0.4,0.6,0.8,1.0,1.2,1.4,1.6,1.8,2.0] & 70 \\

 \bottomrule
\end{tabular}
\caption{Setups for the GMP+SNN models used in the force training example}
\label{tab:forcetrainingSetup}
\end{table*}

\newpage
\subsection{QM9 Example}

\subsubsection{Per-element Bias}

A per-element bias is added to the SNN model to improve performance. Conceptually, this is equivalent to fitting to formation energies rather than absolute energies. The total energy of an atom is the model predicted energy plus the per-element bias of the specific atom type. To determine the bias, a linear model is applied. The number of atoms for each element types are counted for all the images in the training set, and they are the independent variable. The corresponding energies for each system is the dependent variable. For example, the per-element bias of the trials with 100K training images is shown below in Table \ref{tab:QM9Bias}:

\begin{table*}[!h]
\centering
\begin{tabular}{c c }
\toprule
Atom Type & Per-element Bias (meV) \\
\midrule
H   &   -2795.2721\\
C   &   -6217.7719\\
N   &   -4552.4431\\
O   &   -4432.6761\\
F   &   -4075.4931\\
 \bottomrule
\end{tabular}
\caption{Per-element Bias of each atom type found by the linear model, for the 100K training set}
\label{tab:QM9Bias}
\end{table*}

\subsubsection{Training Procedure}
With the per-element bias determined, GMP+SNN models are fitted to the atomization energy minus the per-element biases. The model setups are given in Table \ref{tab:QM9Setup}. The cutoff distance is always 15 \AA{}, so that the largest radial probe takes a negligible value of $8\times10^-4$ at the cutoff. The models are trained for 12,000 epochs with learning rate decrease by factor of 2 every 2,000 epochs, from $1^{-2}$ to $3^{-4}$. The batch size is set to be 32 images.

\begin{table*}
\centering
\begin{tabular}{c c c c}
\toprule
Model & Sigmas & $N_{feature}$ & $N_{parameters}$\\ \midrule
GMP(10,1)+SNN(32,32,32) & linspace(0.02, 2.0, 10, endpoint=True) & 20 & 3009 \\ \midrule
GMP(30,2)+SNN(128,64,64) & linspace(0.02, 2.0, 30, endpoint=True) & 90 & 24641 \\ \midrule
GMP(50,4)+SNN(256,128,64) & linspace(0.02, 2.0, 50, endpoint=True)  & 250 & 106369\\ 
\midrule
GMP(70,6)+SNN(512,256,128,64) & linspace(0.02, 2.0, 70, endpoint=True)  & 490 & 425857\\

 \bottomrule
\end{tabular}
\caption{Setups for the GMP+SNN models used in the QM9 examples. Cutoff distance is always 15 \AA{}. $N_{parameters}$ is the number of trainable parameters of the neural network model.}
\label{tab:QM9Setup}
\end{table*}

\subsubsection{Transfer Learning to New Element}
For this example, we used the same procedure as above, with the caveat that the per-element biases are not fitted using a linear model, but directly pulled from the 100k molecule trial. For more detail please refer to the test scripts.

\begin{figure}[!ht]
	\centering
    \includegraphics[width=0.6\linewidth]{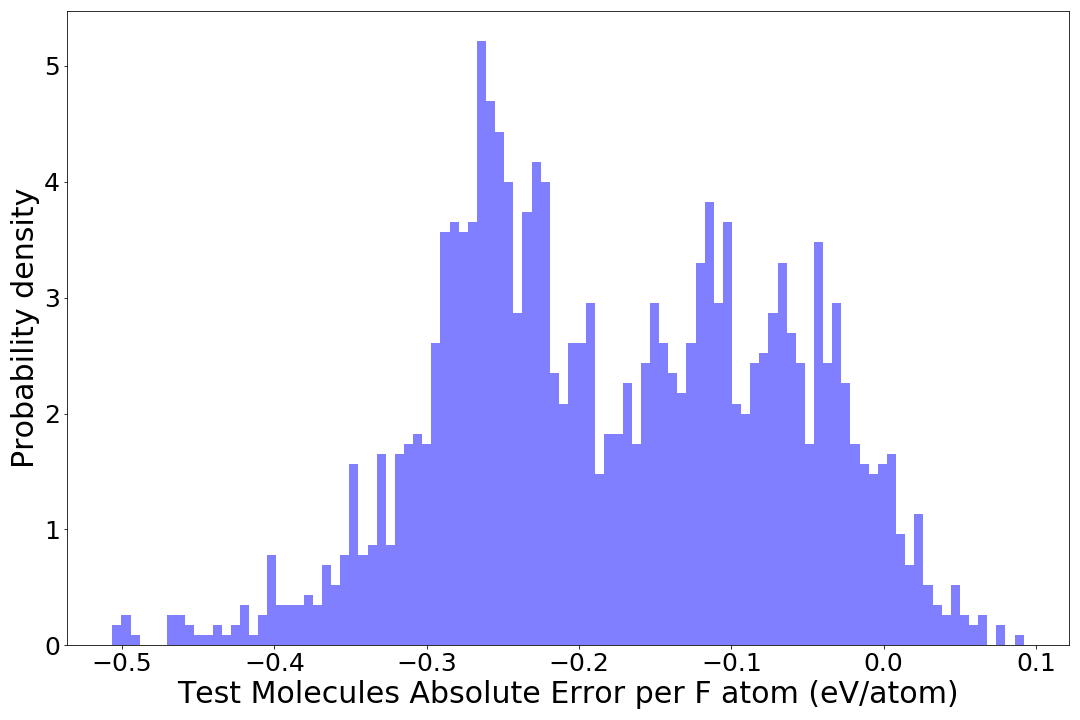}
    \caption{Error distribution of F-containing molecules per F atom.}
    \label{fig:QM9TransferPerF}
\end{figure}

\begin{figure*}[!ht]
	\centering
    \begin{subfigure}{0.48\textwidth}
		\centering
        \includegraphics[width=\linewidth]{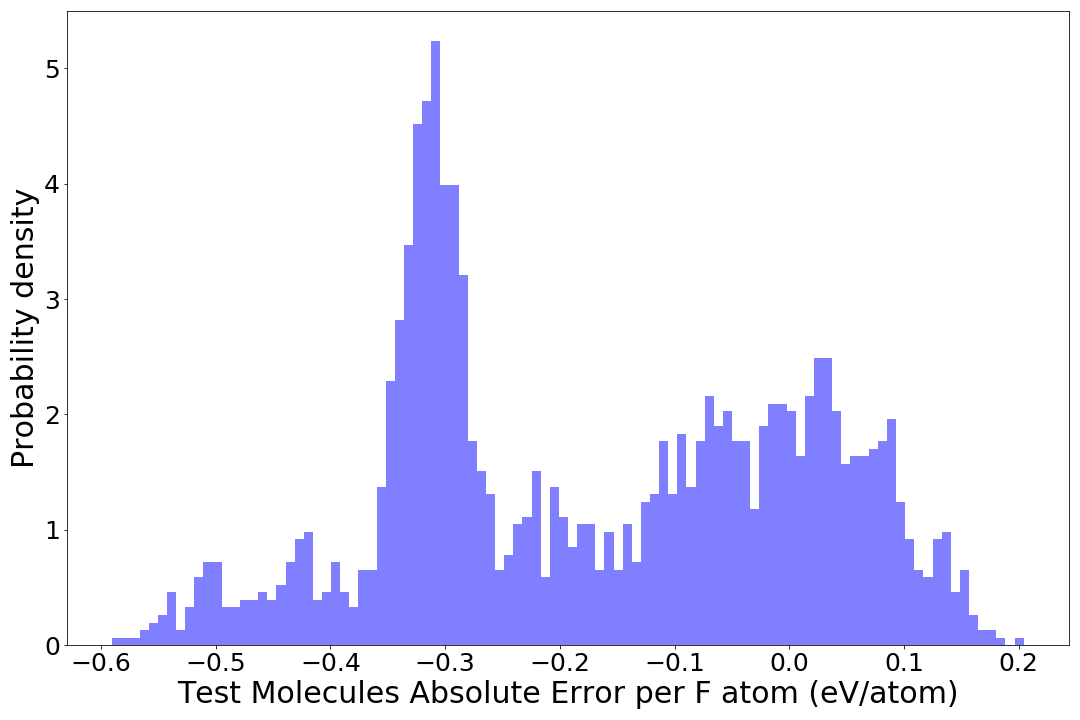}
        \caption{}
        \label{fig:QM9TransferDifferentSeeds0}
	\end{subfigure}\hspace{0.005\textwidth}
    \begin{subfigure}{0.48\textwidth}
		\centering
        \includegraphics[width=\linewidth]{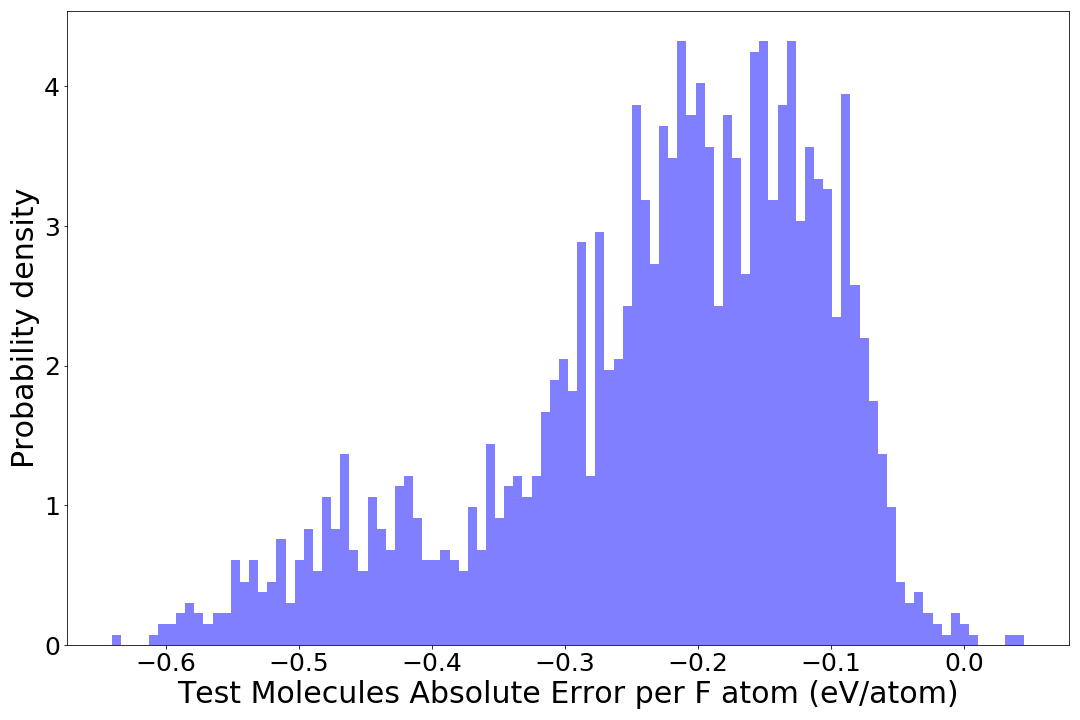}
        \caption{}
        \label{fig:QM9TransferDifferentSeeds1}
	\end{subfigure}\hspace{0.005\textwidth}%
	    \begin{subfigure}{0.48\textwidth}
		\centering
        \includegraphics[width=\linewidth]{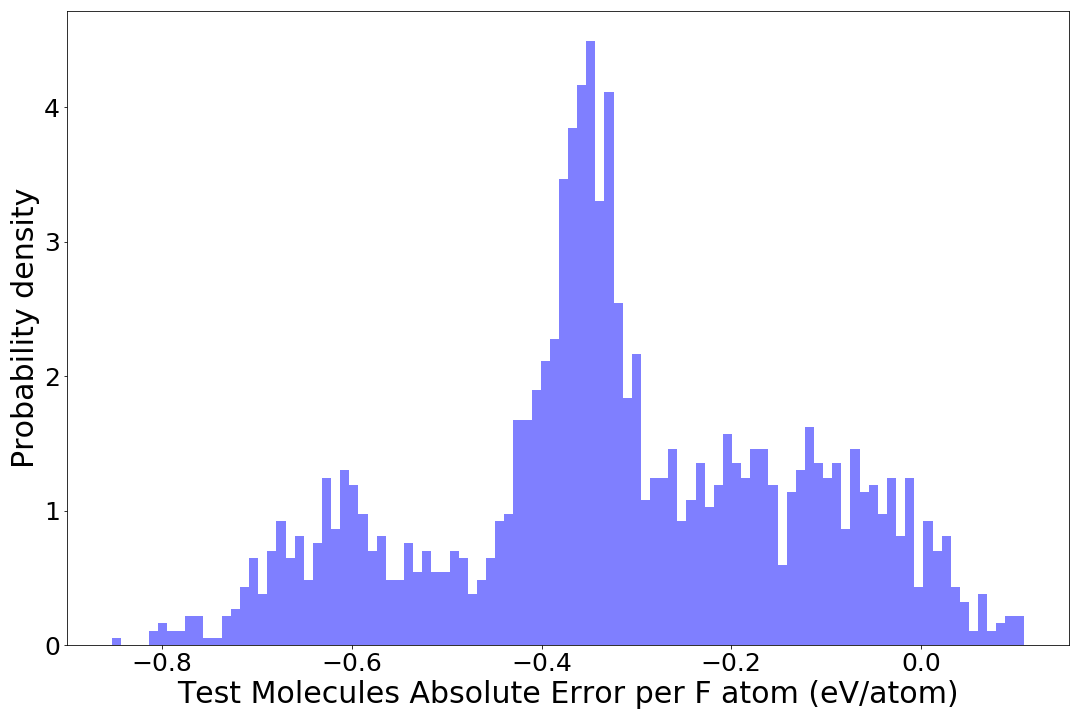}
        \caption{}
        \label{fig:QM9TransferDifferentSeeds2}
	\end{subfigure}\hspace{0.005\textwidth}
    \begin{subfigure}{0.48\textwidth}
		\centering
        \includegraphics[width=\linewidth]{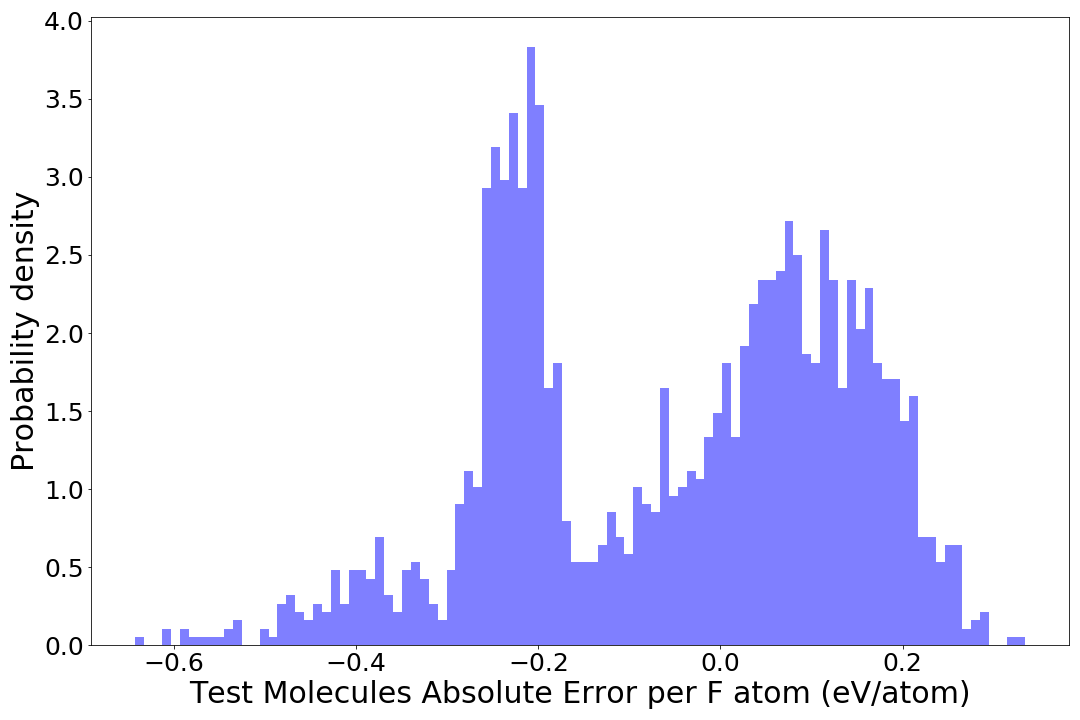}
        \caption{}
        \label{fig:QM9TransferDifferentSeeds3}
	\end{subfigure}\hspace{0.005\textwidth}%
    \caption{Error distribution of prediction for F-containing molecules with different random seed for the model trained on molecules without F.}
    \label{fig:QM9TransferDifferentSeeds}
\end{figure*}

Shown in Figure \ref{fig:QM9TransferPerF} is the error distribution of the system containing F atoms in the basis of error per F atom. Shown in Figure \ref{fig:QM9TransferDifferentSeeds} are the error distributions from different trials with different random seeds.

\newpage

\begin{table*}
\centering
\begin{tabular}{c c c c}
\toprule
Model & Sigmas & $N_{feature}$ & $N_{parameters}$\\ \midrule
GMP(30,2)+SNN(128,64,64) & linspace(0.02, 2.0, 30, endpoint=True) & 90 & 24641 \\ \midrule
GMP(50,4)+SNN(256,128,64) & linspace(0.02, 2.0, 50, endpoint=True)  & 250 & 106369\\ 
\midrule
GMP(70,6)+SNN(512,256,128,64) & linspace(0.02, 2.0, 70, endpoint=True)  & 490 & 425857\\
\midrule
GMP(90,8)+SNN(1024,512,128,64) & linspace(0.02, 2.0, 90, endpoint=True) & 910 & 1432705 \\

 \bottomrule
\end{tabular}
\caption{Setups for the GMP+SNN models used in the OC20 examples. Cutoff distance is always 15 \AA{}. $N_{parameters}$ is the number of trainable parameters (weights) of the neural network model.}
\label{tab:OC20Setups}
\end{table*}

\end{document}